\documentclass{article}

     \PassOptionsToPackage{numbers, compress}{natbib}

\usepackage[preprint]{neurips_2023}
\usepackage{graphicx}

\usepackage{layout}
\usepackage{multirow} 
\usepackage[utf8]{inputenc} 
\usepackage[T1]{fontenc}    
\usepackage[pagebackref=true,breaklinks,colorlinks,citecolor=blue,linkcolor=blue,urlcolor=blue,hypertexnames=false]{hyperref}
\usepackage{url}            
\usepackage{booktabs}       
\usepackage{amsfonts}       
\usepackage{nicefrac}       
\usepackage{microtype}      
\usepackage{xcolor}         

\title{Learning Frame-Wise Emotion Intensity for Audio-Driven Talking-Head Generation}

\author{%
   Jingyi Xu \\
   Stony Brook University  \\
   \And
   Hieu Le \\
  EPFL \\
   \And
  Zhixin Shu \\
   Adobe Research \\
   \And
   Yang Wang \\
   Waymo \\
   \AND
   Yi-Hsuan Tsai \\
   Google \\
    \And
   Dimitris Samaras \\
   Stony Brook University \\
}

\begin{document}
\newcommand{\hl}[1]{{\color{orange} #1}}

\maketitle

\begin{abstract}
Human emotional expression is inherently dynamic, complex, and fluid, characterized by smooth transitions in intensity throughout verbal communication. However, the modeling of such intensity fluctuations has been largely overlooked by previous audio-driven talking-head generation methods, which often results in static emotional outputs.
In this paper, we explore how emotion intensity fluctuates during speech, proposing a method for capturing and generating these subtle shifts for talking-head generation. 
Specifically, we develop a talking-head framework that is capable of generating a variety of emotions with precise control over intensity levels. This is achieved by learning a continuous emotion latent space, where emotion types are encoded within latent orientations and emotion intensity is reflected in latent norms. In addition, to capture the dynamic intensity fluctuations, we adopt an audio-to-intensity predictor by considering the speaking tone that reflects the intensity.
The training signals for this predictor are obtained through our emotion-agnostic intensity pseudo-labeling method without the need of frame-wise intensity labeling. Extensive experiments and analyses validate the effectiveness of our proposed method in accurately capturing and reproducing emotion intensity fluctuations in talking-head generation, thereby significantly enhancing the expressiveness and realism of the generated outputs.

\end{abstract}

\section{Introduction}

Audio-driven talking-head synthesis \cite{das2020motion,EAT,sad-talker,pc-avs,wav2lip} has drawn increasing attention due to its wide range of applications in various fields such as virtual reality, digital humans \cite{Ad-nerf}, and assistive technologies. Recently, many works
have focused on synchronizing lip movements with speech content \cite{wav2lip, chen2018lip,chen2020rhythmic,song2018rnn}.
However, in addition to speech, humans also convey intentions through emotional expressions. Hence, developing emotionally expressive talking-heads is crucial to enhance the fidelity of these systems for real-world applications. 

Unlike static displays of emotion, our emotional expressions are inherently dynamic and fluid with smooth transition in terms of emotion intensity levels. Imagine a scenario where someone receives unexpected good news. Initially, they may smile slightly as they process it. As they fully comprehend, their smile might widen and their eyes might brighten, reflecting growing excitement and happiness. This progression illustrates the dynamic nature of human expressions with smooth shifts in emotion intensity levels, unlike static displays such as a fixed smile.

While there have been previous efforts to generate emotion-aware talking-heads \cite{EAT,SPACE,mm-esl,evp}, the natural fluctuations in intensity levels have been largely overlooked. Addressing this requires the model to 1) infer smooth emotion intensity flows based on audio cues and 2) precisely generate facial expressions that correspond to the inferred intensity levels.
However, capturing these intensity fluctuations can be challenging in existing emotional-aware talking-head methods: 1) the model would rely on frame-wise intensity annotations, which are difficult to obtain in practice; 2) there should be mechanisms to control the generated intensity precisely, instead of homogeneous outputs.

In this paper, we propose a framework for talking-head generation that effectively tackles the two challenges as mentioned above, resulting in talking-head videos with natural intensity flows. In particular, we first introduce a method to obtain audio-synchronized intensity fluctuations without any frame-level annotations. This is achieved through an emotion-agnostic intensity pseudo-labeling method that accurately quantifies the intensity of any given frame based on keypoint distance. An audio-to-intensity predictor is then employed to predict these frame-wise pseudo intensities based on the speaking tone, as the speaking tone often reflects emotional fluctuations.

Furthermore, our goal is to generate talking-heads with emotions that correspond accurately to the inferred intensity levels. This task is particularly challenging because varying intensities often manifest as subtle and smooth differences, which are difficult to capture. We draw inspiration from the ``wheel of emotion'' \cite{emowheel}, where emotions are arranged on a disk with similar emotions positioned close to one another, and intensity is represented by the distance from the center. Instead of using a disk, we propose constructing a continuous and unified latent space that naturally represents all types of emotions at various intensity levels. In this space, the neutral emotion is placed at the origin, with similar emotions located nearby. The emotion type is encoded in the latent orientation, while the intensity is represented by the latent norm. This design enables smooth transitions between emotional intensities and types, enhancing the efficiency of the learning process. Quantitative results on MEAD \cite{mead} and LRW \cite{lrw}, along with qualitative analyses, demonstrate the effectiveness of this method in generating diverse expressions with precise control over intensity levels.

Our main contributions can be summarized as follows:
\begin{itemize}
    \item We present a talking-head generation framework that considers the emotion intensity for producing smooth transitions across facial expressions.

    \item We introduce an audio-to-intensity predictor to capture the dynamics of emotion intensity in audio input, without the need of having frame-wise ground truth labels for intensity.

    \item We propose an approach for reorganizing the latent space of emotions in talking-head generation, enabling both diverse emotional expressions and precise control over intensity levels.

\end{itemize}
\section{Related Work}

\noindent{\textbf{Audio-Driven Talking Head Generation.}}
Early works on audio-driven talking-head generation mainly focus on producing accurate lip motion~\cite{wav2lip, chen2018lip,chen2020rhythmic,song2018rnn,zhou2019disentangle}. Wav2Lip~\cite{wav2lip}, for example, leverages audio signals to synchronize lip movements within a provided video sequence. Subsequent advancements have led to techniques for generating full-head animations~\cite{makeittalk,ren2021pirenderer,wu2021imitating,zhang2021flow,zhang2021facial,yin2022styleheat}. Audio2Head~\cite{audio2head} employs keypoint-based dense motion fields to manipulate facial images for enhanced realism.
Recently, transformer-based architecture has been adopted for this task due to its ability to capture the long-term context of audio inputs and generate complete sequence-level representations.
AVCT~\cite{avct} designs an audio-visual correlation transformer
for generating talking-head videos. EAT~\cite{EAT} utilizes an audio-to-expression transformer to map audio sequences to enhanced 3D latent keypoints.
In this paper, different from the above-mentioned approaches, we focus on another important aspect of talking-head generation, i.e., precise modeling of emotion intensity and fluctuations, thereby enriching the naturalness of generated results.

\noindent{\textbf{Emotion-Aware Talking Head Generation.}}
Emotional talking-head generation has recently attracted growing interest within the field \cite{EAT,SPACE,mm-esl,evp,emmn,mead,emotalk,GC-AVT,sinha2022emotion}. 
Wang \textit{et al.} release MEAD \cite{mead}, a high-quality talking-head video dataset annotated with emotion categories and intensity. Gan \textit{et al.} \cite{EAT} use this dataset to fine-tune a pre-trained emotion-agnostic talking-head transformer for emotion-aware talking-head generation. 
SPACE \cite{SPACE} leverages facial landmarks and latent keypoints as intermediate face representations, facilitating talking-head generation with control over emotion type and intensity level.
However, these methods encode emotion labels using discrete one-hot vectors, which restricts the diversity of the generated expressions.
Moreover, MM-ESL \cite{mm-esl} constructs a unified feature space for various emotion types and controls the expression simply with a scalar parameter.
In contrast, we propose to model the intensity directly from the audio input in an emotion-agnostic manner, so our learned intensity can be flexibly integrated in our emotion latent space without influencing the expression generation process.
%

\noindent{\textbf{Latent-Based Face Editing.}}
Manipulating latent space is a common practice for face-editing tasks \cite{attgan,fadernet,genegan,elegant,indomaingan, styleflow, latent_transformer,latent_directions_gan,neuralodes}. FaderNet \cite{fadernet} disentangles different attributes in the latent space of auto-encoder via adversarial training.
InterfaceGAN \cite{interfacegan} focuses on understanding the semantics within the latent space of generative adversarial networks (GANs) \cite{gan} through subspace projection. L2M-GAN \cite{l2m-gan} showcases its ability for both local and global face attribute editing through latent space factorization. In this work, we present an intuitive approach for reorganizing the latent space of emotions in talking-head generation, which allows for a wide range of emotional expressions and offers precise control over intensity levels.

\begin{figure}[!t]
\begin{center}
\hspace{-2mm}
\includegraphics[width=0.92\columnwidth]{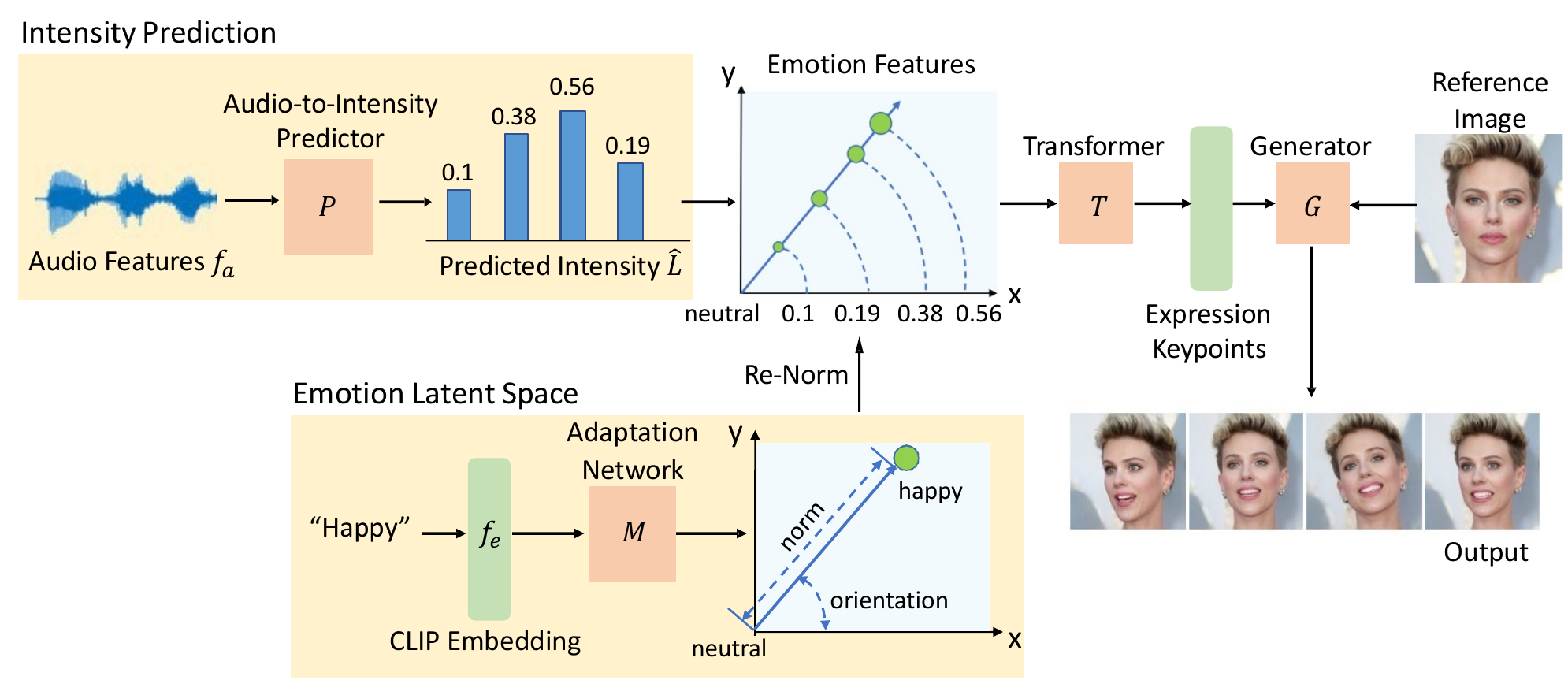} \vspace{1mm}
\caption{
An overview of our proposed method. 
To generate talking-heads with fluid intensity transitions, we first infer these fluctuations from audio inputs using an audio-to-intensity predictor $\mathbf{P}$. The training targets for this predictor are obtained through our emotion-agnostic intensity pseudo-labeling method. 
Then, for a specified driving emotion $e$ (e.g., \textit{happy}), we map it onto our proposed emotion latent space to get $M(f_e)$ and adjust its norm based on the inferred intensity $\hat{L}$. These resulting intensity-aware emotion features $M^r(f_e)$ then serve as guiding signals for a transformer-based talking-head generation model, enabling the generation of lifelike talking-head videos.
} \vspace{-0mm}
\label{fig:pipeline}
\end{center}
\end{figure}

\section{Our Proposed Method}
\label{method}
In this section, we introduce our proposed method for generating talking-heads with frame-wise emotion intensity. Figure \ref{fig:pipeline} shows an overview of the pipeline. Specifically, to generate talking-heads with fluid intensity transitions, we first infer these fluctuations from audio input using an audio-to-intensity predictor. 
Then, for a specified driving emotion (e.g., \textit{happy}), we map it onto our proposed emotion latent space and adjust its norm based on the inferred intensity fluctuations. These resulting intensity-aware emotion features then serve as signals for a transformer-based talking-head generation model, producing natural talking-head videos. In the following subsections, we will elaborate on the details of our intensity pseudo-labeling method, the audio-to-intensity predictor training, and the reorganization of the intensity-aware emotion latent space.

\subsection{Frame-Wise Intensity Modeling}

In order to learn the intensity directly from the audio input, we propose a pseudo-labeling method to avoid the frame-wise labeling requirement, which is challenging to achieve in practice. In the following, we describe more details about our pseudo-labeling scheme and how we train the audio-to-intensity model.

\noindent{\textbf{Emotion-agnostic intensity pseudo-labeling.}}
To model the dynamic fluctuations in intensity during natural speech, we first introduce an emotion-agnostic intensity pseudo-labeling approach. This method quantifies the intensity level of any given frame.
Specifically, given a talking-head frame $i$ and the corresponding expression keypoints $K_i$, we extract the expression keypoints $K_i^N$ from the neutral face of the same person (see supplementary materials for more details). 
The pseudo intensity $L_i$ is then computed as the sum of absolute difference between $K_i$ and $K_i^N$, i.e., $L_i = \sum \left| K_i - K_i^N \right|$. Essentially, a greater deviation between the given expression and the neutral facial expression indicates a higher intensity level, while a lower deviation suggests a lower intensity level.

\noindent{\textbf{Audio-to-intensity predictor.}}
To further estimate the intensity fluctuations from audio input during inference, we introduce a variational auto-encoder based model for audio-to-intensity prediction. This model predicts the pseudo-labeled emotion intensity conditioned on audio input, as the speaking tone often reflects the emotion intensity when people talk. To obtain audio representations, we utilize HuBERT \cite{hubert}, a state-of-the-art automatic speech recognition model to extract audio features $f_a$ from the input waveform. Throughout training, the intensity predictor ($\mathbf{P}$) is trained to predict the entire sequence of pseudo intensity using an encoder-decoder structure. The training loss of this predictor can be written as follows:
\begin{equation}\label{eq:vae_loss}
    L_{P}(\phi, \theta, \epsilon) = -\textnormal{E}_{q_\phi(z|L,a)}[\textnormal{log}p_\theta(L|z, a)] + \textnormal{KL}(q_\phi(z|L,a)|p_\epsilon(z|a)),
\end{equation}
where $\phi, \theta, \epsilon$ denote the model parameters of the encoder, decoder and the VAE prior, respectively. $z$ denotes the latent encoding of the intensity.
A normalizing flow is employed to establish a time-related distribution as the prior, following the approach in \cite{GeneFace}.
We provide further details about the architecture of the encoder and decoder in the supplementary materials.

\subsection{Emotional Talking-Head Model Training}

We present our emotion-aware talking-head generation model, which leverages the predicted intensity fluctuations as input to generate corresponding facial expressions. To this end, we introduce an emotion latent space in which the latent features are used to guide a transformer-based model to precisely generate the predicted emotion intensity.

\subsubsection{Emotion Latent Space Reorganization}

To enable emotional talking-head generation, we propose to learn a continuous emotion latent space where the latent embeddings serve as guidance to steer an audio-to-expression transformer in generating different emotional expressions. 
Specifically, we expect this latent space to 1) accommodate features of diverse emotion categories, and 2) encode corresponding intensity levels. Furthermore, we aim for disentanglement between the encoded emotion category and intensity, thereby allowing for flexible control over both.
To achieve this, we reorganize the latent space such that emotion category information is encoded in the latent orientation (see Figure \ref{fig:pipeline}), while intensity information is encoded in the latent norm. For neutral emotion, we consider the intensity to be zero and position it at the origin of the latent space. Our intuition is that the intensity level should positively correlate with the deviation from a neutral expression. Through our reorganization method, this deviation is intrinsically encoded in the distance from the origin point in the latent space, i.e., latent norm.

\noindent{\textbf{Intensity-aware emotion features.}}
For an input video with emotion label $e$ and intensity label $l$, we extract the corresponding text embedding $f_e$ from the CLIP \cite{clip} text encoder. An adaptation network $M$ is then adopted to map $f_e$ to our proposed emotion latent space. In the case where $e$ represents neutral emotion, we minimize the L2-norm of the emotion feature $M(f_e)$ to enforce it to center around the origin. Otherwise, for non-neutral emotions, we re-scale $M(f_e)$ to $M^r(f_e)$ such that $\|M^r(f_e)\| = t(l)$, where $t(l)$ is positively correlated with $l$. This ensures that the features of the same intensity will be mapped to the same hypersphere, with norms representing intensity levels. Meanwhile, the location on this hypersphere is determined by the orientation of $M(f_e)$, which is shared within the same emotion type.

\subsubsection{Training Objectives}
We input the re-scaled emotion features to an audio-to-expression transformer $\mathbf{T}$ as a guidance to generate emotional expressions. Specifically, this guidance is integrated as an additional input token of the transformer layer following \cite{EAT}. The transformer produces a set of intensity-aware 3D expression latent keypoints, which are then used by a talking-head generator $\mathbf{G}$ to generate video frames.
The final loss for training this talking-head generation system is calculated as follow:
\begin{equation}\label{eq:loss}
    L = \lambda_{exp} * L_{exp} + \lambda_{rec} * L_{rec} + \lambda_{sync} * L_{sync} + \lambda_{norm} * L_{norm},
\end{equation}
where $\lambda_{exp}$, $\lambda_{sync}$, $\lambda_{rec}$, $\lambda_{norm}$ are hyper-parameters that re-weight the corresponding loss term. $L_{exp}$ represents the mean square error between the ground-truth expression keypoints and the corresponding keypoints predicted by the transformer. 
$L_{rec}$ is the L1 reconstruction loss between the generated frame and the input frame within the facial region.
$L_{sync}$ is the lip synchronization loss introduced in Wav2Lip \cite{wav2lip}.
$L_{norm}$ denotes the L2-norm of the emotion feature, which only applies for neutral emotion.



\noindent{\textbf{Inference procedure for talking-head generation.}}
Our goal is to generate talking-head videos with emotion intensity fluctuations that correspond to changes in audio inputs. In particular, given an input audio and a driving emotion in a text form, we first infer the emotion intensity $\hat{L}$ using our trained audio-to-intensity predictor $\mathbf{P}$, where the predicted intensity $\hat{L}$ is represented as a sequence of scalars. 
To generate talking-head videos, we map the CLIP embedding of the driving emotion to our proposed emotion latent space. This emotion embedding is then re-normalized according to the predicted intensity $\hat{L}$, resulting a sequence of embeddings with the same latent direction but varying latent norms. With this sequence of intensity-aware emotion embeddings, we can then leverage them to guide the transformer model in generating talking-head videos with frame-wise intensity fluctuations.

\section{Experimental Results}
\label{experiments}

\subsection{Experimental Setup}
\label{sec:setup}
\noindent{\textbf{Implementation details.}}
Following \cite{EAT}, we first pre-train the transformer-based talking-head generation model to produce emotion-agnostic talking-heads. In particular, we sample the input videos at 25 FPS, and crop and resize them to 256 $\times$ 256. The input audios are down-sampled to 16kHz and are transformed into mel-spectrograms, with the window length and hop length set to 640. 
We use the pre-trained keypoint detector in \cite{reposnet} to extract canonical keypoints from source images and expression/head pose sequences from driving videos. 
Following \cite{EAT}, the values of the loss weights are set as: $\lambda_{exp} = 100, \lambda_{rec} = 10, \lambda_{sync} = 10, \lambda_{norm} = 0.1$. For videos at three different intensity levels (levels 1, 2, 3), we re-normalize the emotion features to 5, 15, and 30, respectively. 
For training the audio-to-intensity predictor, we process the input audio with a pre-trained HuBERT model \cite{hubert}. We design the encoder and decoder as dilated convolutional networks following \cite{GeneFace}.

\noindent{\textbf{Datasets and baselines.}}
We use Voxceleb2 \cite{voxceleb2} to train our audio-to-intensity predictor and the talking-head generation model. For emotion-aware generation, we fine-tine the pre-trained talking-head model on MEAD \cite{mead}, a high-quality emotional video set with 8 kinds of emotions. To ensure fair comparisons, we split the MEAD dataset into training and testing sets based on identity, using the same test identities as \cite{eamm}. To show the effectiveness of our method, we compare with one-shot talking-head generation methods on the LRW \cite{lrw} and MEAD \cite{mead} test sets. These methods include ATVG \cite{atvg}, Wav2Lip \cite{wav2lip}, MakeItTalk \cite{makeittalk}, AVCT \cite{avct}, PC-AVS \cite{pc-avs}, EAMM \cite{eamm}, and EAT \cite{EAT}.

\noindent{\textbf{Evaluation metrics.}}
We evaluate the quality of generated videos on multiple metrics that have been widely used in previous studies. Specifically, we use Frechet Inception Distance (FID) \cite{fid} to evaluate the realism of generated frames. To evaluate identity preservation, we calculate the structural similarity index measure (SSIM) and peak signal-tonoise ratio (PSNR) \cite{SSIM} of identity embedding between the source images and the generated frames. To evaluate audio-visual synchronization, we use the confidence score provided by SyncNet \cite{syncnet}. In addition, we use the distance between landmarks of the entire face (F-LMD) \cite{atvg} as a measure of pose and expression accuracy. To assess the emotional accuracy (Emo-Acc) of the generated emotions, we fine-tune the Emotion-Fan \cite{emotionfan} using the training set of MEAD following \cite{EAT} and use it for emotion classification.

\subsection{Comparison with State-of-the-Art Methods}

\subsubsection{Quantitative Results}

Following the setting of EAMM \cite{eamm}, we test our method on the public-available MEAD test set for emotional talking-head generation. Results are summarized in Table~\ref{tab:mead_results}. It can be observed that our method outperforms the previous methods by a large margin in terms of emotion accuracy and video quality, and obtains comparable lip-sync results with EAT. The improvement in emotion accuracy is likely due to our method's effective modeling of emotion intensity integrated with the emotion embedding space, resulting in more discriminative emotion representations.
This enables the transformer to generate more accurate and diverse facial expressions, which further leads to better generation results.

In addition, we test our method on the LRW \cite{lrw} dataset for one-shot emotion-agnostic talking-head generation.
LRW dataset comprises 25K neutral videos. We use the first frame of each video as the reference image to generate videos following \cite{sad-talker}. As shown in Table \ref{tab:lrw_results}, our method achieves better overall video quality.
We note that Wav2Lip and PC-AVS might overfit to the pre-trained SyncNet model since their lip-sync scores are even higher than the ground-truth. Moreover, Wav2Lip only generates mouth parts for better lip synchronization while our method specifically focuses on emotion modeling.

\begin{table*}[!t] 
\centering
\scriptsize
\caption{ Quantitative comparisons with state-of-the-art methods on MEAD \cite{mead}.}
\resizebox{0.85\textwidth}{!}{%
  \begin{tabular}{l|cccccc}
    \toprule
    Method & Emo-Acc $\uparrow$ & SSIM $\uparrow$ & PSNR $\uparrow$ & FID $\downarrow$ &  SyncNet $\uparrow$ & F-LMD $\downarrow$ \\
    \midrule
   {ATVG \cite{atvg}}  & 17.36
 & {0.56} & {17.64} & 99.42 & {1.80}   & 3.74    \\
   {Wav2Lip \cite{wav2lip}}  & 
   17.87  & {0.57} & {19.12} & 67.49 & \textbf{8.97} & 3.71   \\
     MakeItTalk  \cite{makeittalk}  & 15.23 & 0.55 & 18.79 & 51.88 & 4.00  & 5.28   \\
     AVCT  \cite{avct}  & 15.64  & 0.54 & 18.43 & 39.18 & 6.02 & 4.33  \\
    {PC-AVS \cite{pc-avs}}  & 11.88 & {0.61} & {20.60} & 53.04 & {8.60} & 2.70  \\
    EAMM \cite{eamm} & 49.85  &  {0.66} & {20.55} & {22.38} & 6.62 & 2.55   \\
    EAT \cite{EAT} & 75.43 &  {0.68} & {21.75} & {19.69} & 8.28 & 2.47  \\
    Ours & \textbf{83.55} & \textbf{0.83} & \textbf{23.66} & \textbf{16.77} & {8.31} & \textbf{2.40}   \\
    \midrule
    Ground-Truth & 84.37 &  {1.00} & $\infty$ & {0}  & 7.76 & 0.00 \\
    \bottomrule
  \end{tabular}
  } \\ 
  \label{tab:mead_results}
  \vspace{-2mm}
\end{table*}

\begin{table*}[!t] 
\centering
\scriptsize
\caption{Quantitative comparisons with state-of-the-art methods on LRW \cite{lrw}.}
\resizebox{0.7\textwidth}{!}{%
  \begin{tabular}{l|ccccc}
    \toprule
    Method & SSIM $\uparrow$ & PSNR $\uparrow$ & FID $\downarrow$ & SyncNet $\uparrow$ & F-LMD $\downarrow$  \\
    \midrule
   {ATVG \cite{atvg}}  
 & {0.64} & {18.40} & 51.56  & {2.73}  & 3.31  \\
   {Wav2Lip \cite{wav2lip}}  & 
   {0.73} & {22.80} & 7.44  & \textbf{7.59} & 2.47  \\
     MakeItTalk  \cite{makeittalk}  & 0.69 & 21.67 & 3.37 & 3.28 & 2.99 \\
     AVCT  \cite{avct}   & 0.68 & 21.72 & 2.01   & 4.63 &  3.23 \\
    {PC-AVS \cite{pc-avs}}  & {0.72} & {23.32} & 4.64 & {7.36}  & 2.11  \\
    EAMM \cite{eamm} &  {0.71} & {22.34} & {6.44}  & 4.67  & 2.37 \\
    EAT \cite{EAT} &   {0.77} & {24.11} & {3.52}  & 6.22 & 2.08 \\
    Ours &  \textbf{0.85} & \textbf{25.63} & \textbf{1.88}  & {6.20} & \textbf{1.75} \\
    \midrule
    Ground-Truth  &   {1.00} & {$\infty$} & {0}  & 7.06 & 0.00 \\
    \bottomrule
  \end{tabular}
  } \\
\label{tab:lrw_results} \vspace{-4mm}
\end{table*}

\begin{figure}
\vspace{2mm}
\begin{center}
\hspace{-8mm}
\includegraphics[width=0.88\columnwidth]{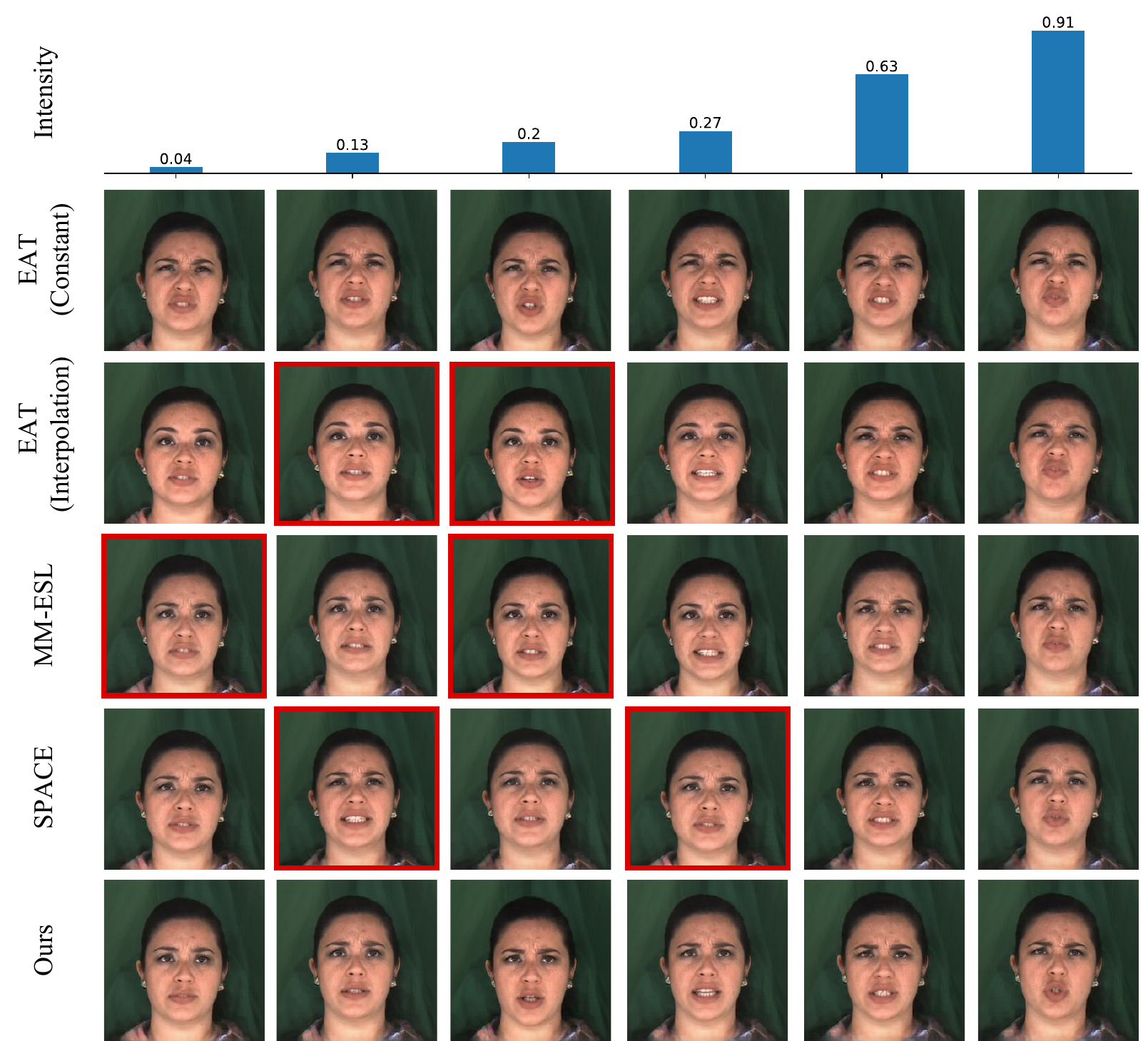} \vspace{1mm}
\vspace{-3mm}
\caption{
We compare our method with several intensity-control methods for audio-driven talking-head generation. The first row represents the predicted intensity from the audio-to-intensity predictor, while the subsequent rows showcase the results of different methods utilizing the predicted intensity. We observe that the intensity of the emotions generated by our method highly correlates with the target intensity, resulting in more diverse and realistic talking-head results. In contrast, the generated intensity from other methods does not consistently align with the target intensity, as shown by the bounded frames in red.
}
\label{fig:intensity_comparison}
\end{center}
\vspace{-6mm}
\end{figure}

\subsubsection{Qualitative Results}
We show visual results of different emotion intensity-control methods in Figure \ref{fig:intensity_comparison}. The first row is the predicted intensity from the audio-to-intensity predictor, which is then used as the target intensity to generate \textit{angry} talking-heads. The second row shows the results of EAT using constant intensity. The generated \textit{angry} emotions are constantly intense throughout the video without any fluctuations.
The results in the third row are from interpolating the emotion embedding of \textit{angry} with \textit{neutral} from EAT.
We show that the model hardly generates meaningful emotions with intensity levels in-between.
MM-ESL \cite{mm-esl} introduces an additional scalar for intensity control, but the generated intensity is not precisely correlated with the target intensity, especially when the values are low (as shown in the first four columns).
SPACE \cite{SPACE} combines the emotion label with its intensity during training. While the generated images exhibit some intensity variation, it is noteworthy that the face generated with intensity ($0.13$) in the second column appears more \textit{angry} compared to faces with higher intensity values, i.e., $0.2$ and $0.27$.
In comparison, our proposed method produces diverse and authentic results with accurate intensity control, leading to more natural and realistic talking-head videos.

\subsection{Analysis}

In this section, we conduct additional experiments and analyses to validate the effectiveness of our method.
The existing emotion-related metric mainly focuses on measuring emotion classification accuracy. 
We further evaluate the quality of the generated emotions from three perspectives: the diversity of facial expressions, the alignment between the audio and generated facial movements, and the reconstruction accuracy of emotion intensity. Specifically, for the diversity, we compute the average feature distance of the generated expression keypoints following \cite{siyao2022bailando}. As for the alignment between audio and generated facial movements, we calculate the average temporal distance between each beat in audio and its closest beat in facial movements as the Beat Align Score following \cite{siyao2022bailando}. Regarding the reconstruction accuracy of emotion intensity, we compute the mean square error between the generated emotion intensity and the target intensity, which is the output of the audio-to-intensity predictor. Diversity and beat alignment mainly assess the naturalness of generated talking-heads, while intensity reconstruction accuracy measures model's intensity controllability.

\subsubsection{Analysis on Emotion Intensity Controllability}

In this section, we conduct a comparative analysis of our method for controlling emotion intensity against other intensity-controlling baselines including 1) interpolating the embedding of target emotion with neutral emotion, 2) concatenating the emotion embedding with an intensity scalar $\mu$, where $\mu \in \{1,2,3\}$
during training \cite{SPACE} and 3) scaling the emotion embedding with the intensity scalar $\mu$ \cite{mm-esl}. We implement these three methods within the framework of EAT \cite{EAT}. As shown in Table \ref{tab:intensity}, our proposed method constantly outperforms other baselines. 
By directly leveraging the property of the reorganized latent space, our method achieves consistent controllability of intensity, resulting in more natural talking-head videos, without the need for interpolating intensities based on pre-computed emotions or introducing additional scalar parameters.
%

\begin{table*}[!t] 
\caption{Analysis on emotion intensity controllability on MEAD.}
\centering
\resizebox{0.78\textwidth}{!}{%
  \begin{tabular}{l|c|ccc}
    \toprule
   {Method}  & {Intensity-control} &  Diversity  $\uparrow$ & Beat Align $\uparrow$ & Intensity-L2 $\downarrow$  \\
    \midrule
   {EAT \cite{EAT}}  & Interpolation
 & {7.152} & 0.3081 & 2.6320    \\
    SPACE \cite{SPACE} & Concatenation  &  {6.861} & {0.3020} & {3.2358} \\
    MM-ESL \cite{mm-esl}  & 
   Scaling & {7.495} & {0.3023} & 3.5265  \\
    Ours & Re-Normalization & \textbf{7.714} & \textbf{0.3131} & \textbf{2.5858}  \\
    \bottomrule
  \end{tabular}
  }
  \vspace{-1mm}
  \label{tab:intensity} 
\end{table*}

\begin{table}[!t] 
\caption{Frame-wise intensity v.s. constant intensity on MEAD.}
\vspace{-2mm}
\centering
\resizebox{0.88\textwidth}{!}{%
  \begin{tabular}{l|c|c|ccc}
    \toprule
   {Method}  & Norm & {Intensity} &  Diversity  $\uparrow$ & Beat Align $\uparrow$ & Intensity-L2 $\downarrow$  \\
    \midrule
   {EAT}  & - & Constant & {6.752} & 0.3023 & 3.5534    \\
    Ours & Constant & Constant  &  {7.208} & {0.3024} & {3.5521} \\
    {Ours}  & Intensity-correlated &
   Constant & {7.251} & {0.3040} & 3.5527  \\
    Ours & Intensity-correlated  &Frame-wise & \textbf{7.714} & \textbf{0.3131} & \textbf{2.5858}  \\
    \bottomrule
  \end{tabular} 
  }
  \label{tab:constant_intensity} 
\end{table}

\begin{figure}[!t]
\begin{center}
\includegraphics[width=1\columnwidth]{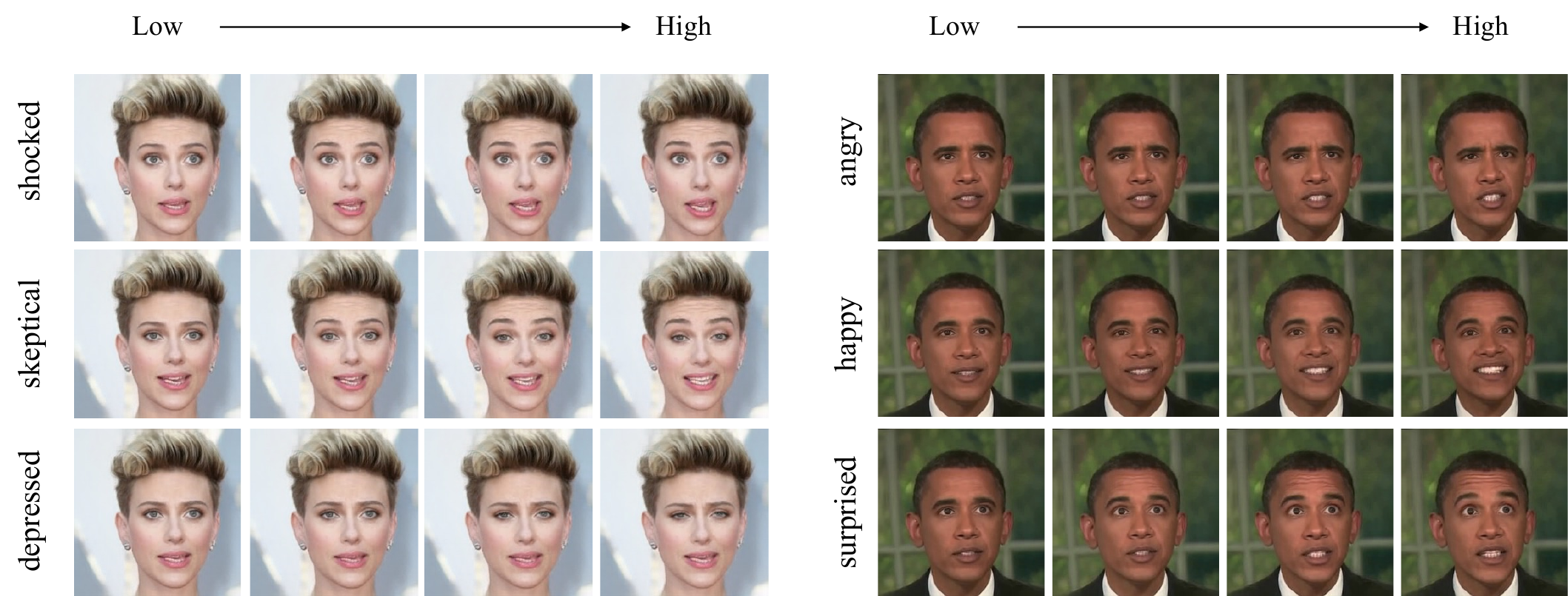} \vspace{1mm}
\vspace{-6mm}
\caption{
Generalization to unseen emotions and identities. 
} \vspace{-3mm}
\label{fig:unseen}
\end{center}
\end{figure}

\subsubsection{Frame-Wise Intensity v.s. Constant Intensity}

In this section, we compare our proposed frame-wise intensity modeling method with simply using constant intensity throughout the talking-head video. Apart from the EAT baseline, we include a variant of our method in comparisons: instead of enforcing intensity-correlated norms during training, this variant employs a constant norm regardless of the intensity level. Results are summarized in Table \ref{tab:constant_intensity}. As shown, using frame-wise intensity yields much lower intensity reconstruction errors compared with using constant intensity. Given that the target intensity is inferred from the audio, higher reconstruction accuracy indicates better alignments between audio signals and facial movements. Moreover, applying frame-wise intensity significantly enhances the diversity of generated expressions, leading to overall more realistic talking-head results.

\subsubsection{Generalization to Unseen Emotions and Identities}
in Figure \ref{fig:unseen}, our method can generalize to unseen emotions and identities. Despite being trained only on eight basic emotions, the model generates accurate novel emotions such as \textit{shocked}, \textit{skeptical} and \textit{depressed}.
This is attributed to the rich semantic knowledge in the CLIP space, allowing the model to effectively generalize to unseen emotions located within similar emotion domains. Remarkably, our method also allows for precise control over the intensity levels of unseen emotions. This further suggests that within our learned latent space, emotion type and intensity level are learned effectively.

\subsubsection{Analysis on Intensity Pseudo-labeling}
\begin{figure}[!t]
\begin{center}
\includegraphics[width=0.88\columnwidth]{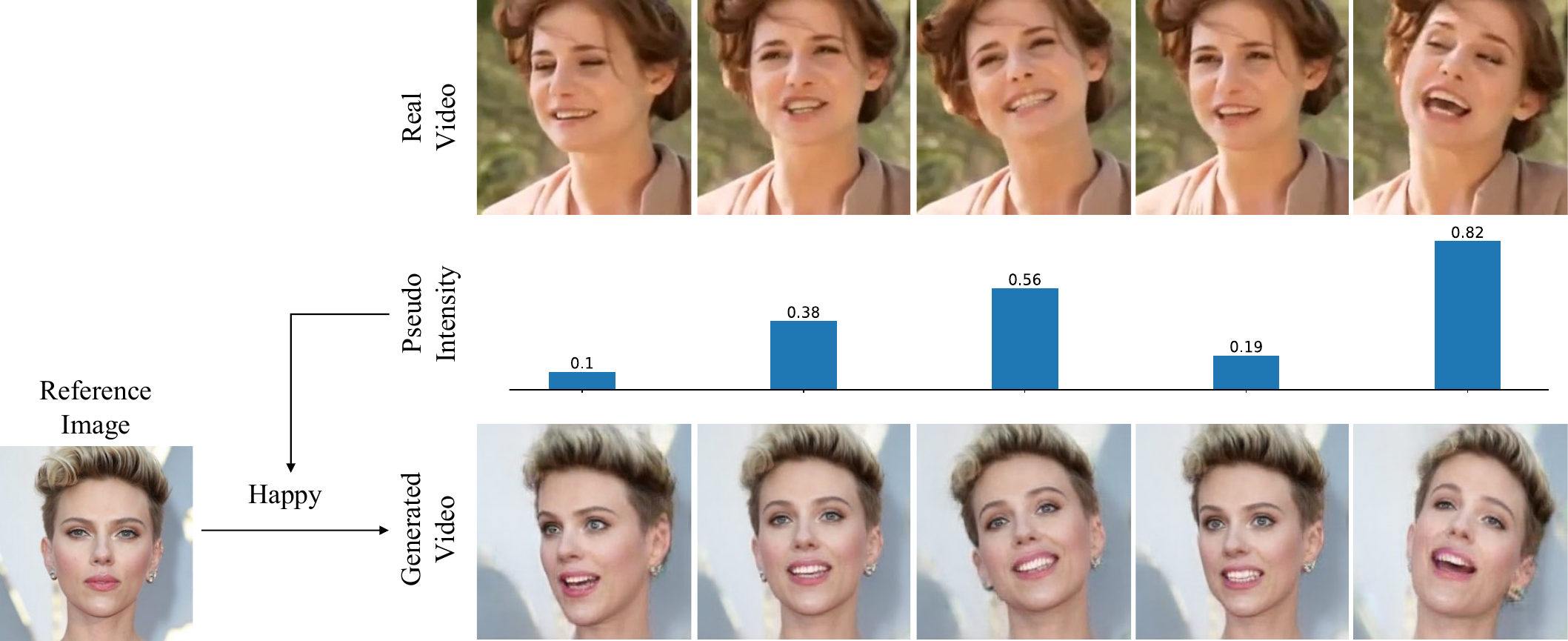}
\vspace{-2mm}
\caption{We show a few frames from a real video, 
the inferred pseudo intensity and the corresponding generated video using the inferred intensity. Our pseudo-labeling method accurately captures the emotion intensity from real talking-heads.
}
\label{fig:transfer}
\end{center}
\vspace{-5mm}
\end{figure}

Due to the lack of frame-wise intensity ground truths, we conduct a qualitative experiment to validate the correctness of our intensity pseudo-labeling method. Specifically, we sample five frames from a real talking-head video and apply our method to compute the pseudo-intensity for each frame. As shown in Figure \ref{fig:transfer}, the computed pseudo-intensity reliably reflects the emotion intensity of each real frame: the \textit{slightly happy} face (in the first column) is assigned a value of $0.1$; while the \textit{extremely happy} face (in the last column) is assigned a value of $0.82$. We further use the pseudo-intensity as the target intensity to generate talking-heads with \textit{happy} emotion (as shown in the third row). The generated emotion intensity precisely corresponds to the target intensity, even under extreme pose variations.

\begin{table*}[!t]
 \caption{User studies on the quality of generated talking-head results.}
\centering
\resizebox{0.77\textwidth}{!}{%
  \begin{tabular}{l|ccc}
    \toprule
   {Method}   &   Intensity Accuracy   & Expression Diversity  & Overall Naturalness  \\
    \midrule
   {MM-ESL} \cite{mm-esl} & {38.89\%} & 13.54\% & 34.12\%    \\
    SPACE \cite{SPACE} &  {42.59\%} & {15.74\%} & {29.41\%} \\
    Ours & \textbf{69.05\%} & \textbf{72.22\%} & \textbf{67.94\%}  \\
    \bottomrule
  \end{tabular}
  }
  \label{tab:user_study} 
   \vspace{-5mm}
\end{table*}

\subsection{User Study}
We conduct user studies involving 25 participants to assess the emotional talking-head videos generated by our method and two other intensity-control methods. For each method, we generate a total of 30 videos, consisting of 20 videos generated from reference images in the test set of MEAD \cite{mead} and 10 videos generated from in-the-wild reference images. Participants are required to evaluate the generated results based on three aspects: 1) accuracy of emotion intensity control, 2) diversity of the generated emotional expressions, and 3) overall naturalness. Results are summarized in Table \ref{tab:user_study}. Our method is preferred by the participants over other two competing methods by a large margin for all three aspects. With our proposed latent space reorganization approach, emotion intensity can be precisely controlled, resulting in diverse facial expression generation and more realistic talking-head videos overall.

\section{Conclusions}
\label{sec:conclusion-limitation}

This paper explores the dynamics of emotion intensity throughout speech, introducing a new approach to capture and replicate these  variations in talking-head generation. Our method begins with developing a talking-head generation model that is capable of generating a variety of emotions with precise control over intensity levels. To further capture the dynamic intensity fluctuations, we propose inferring audio-synchronized intensity fluctuations from the speaking tone with an audio-to-intensity predictor.  Through extensive experiments and analysis, we demonstrate the effectiveness of our approach in accurately capturing and reproducing emotion intensity fluctuations in talking-head generation. This substantially enhances the expressiveness and authenticity of the generated outputs.

\textbf{Ethics Statement.} With the generated expressions at various intensity levels, the talking-head may reflect more proper expressions aligning with the audio, which could lead to better understandings of talking-head applications perceived by humans. On the other hand, although our generated talking-heads achieve state-of-the-art emotion accuracy based on the emotion classification performance, there can be still cases of less obvious or wrong expressions generated by our model. We will aim to develop techniques to detect such cases and refine talking-head results for the ethical consideration.

\bibliographystyle{plainnat}
\bibliography{main}

\newpage
\section{Supplementary Material}

\begin{figure}[!h]
\begin{center}
\hspace{-0mm}
\includegraphics[width=\columnwidth]{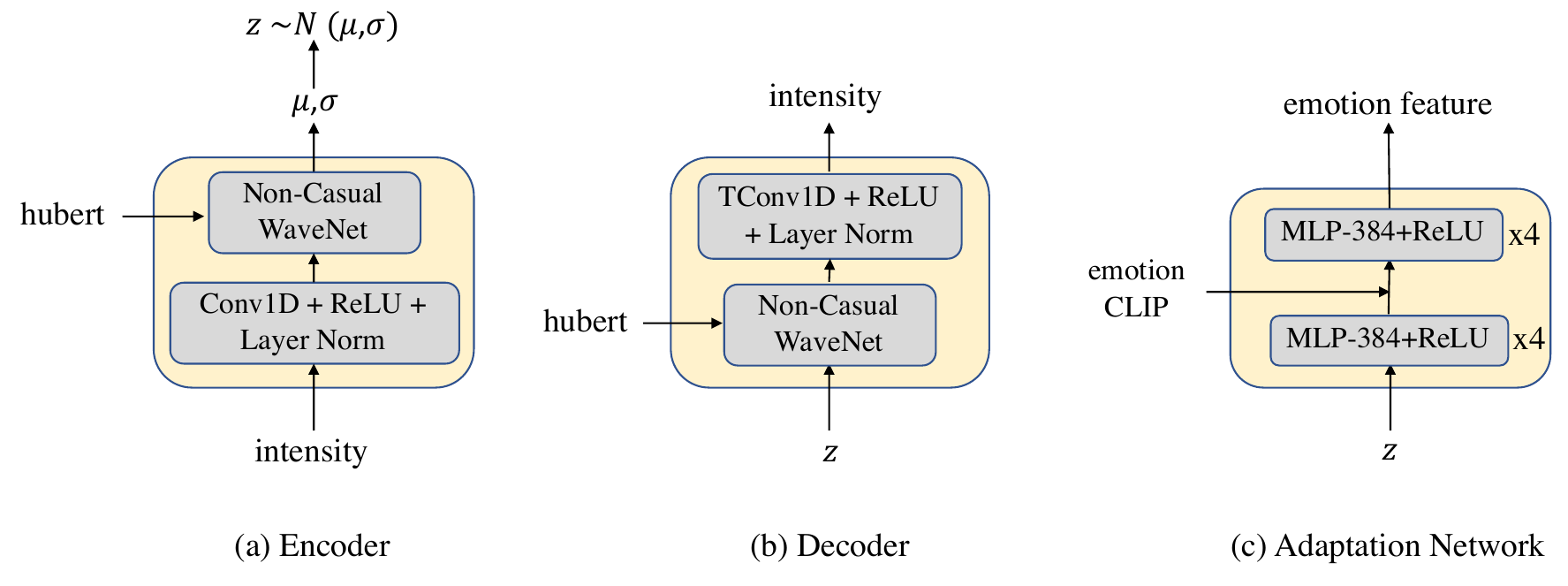} \vspace{1mm}
\caption{ The detailed architecture of the encoder, decoder of the audio-to-intensity predictor and the emotion adaptation network.
} \vspace{-0mm}
\label{fig:architecture}
\end{center}
\end{figure}

\subsection{Details of Model Architectures}

In this section, we provide additional details of the audio-to-intensity predictor and the emotion adaption network. We note that the audio-to-expression transformer is largely based on EAT \cite{EAT}, while the generator is primarily derived from OSFV \cite{reposnet}. Please refer to \cite{EAT} and \cite{reposnet} for more details.

\subsubsection{Audio-to-Intensity Predictor} Our audio-to-intensity predictor consists of an encoder in Figure \ref{fig:architecture}(a), a decoder in Figure \ref{fig:architecture}(b), and a flow-based prior model. The encoder comprises a 1D-convolution, followed by ReLU activation and layer normalization, and then a non-causal WaveNet. The decoder is composed of a non-causal WaveNet and a 1D transposed convolution followed by
ReLU and layer normalization. The prior model is implemented as a normalizing flow, consisting of a 1D-convolution coupling layer and a channel-wise flip operation. We use HuBERT features as the audio condition of these three modules.

\subsubsection{Emotion Adaptation Network} Our emotion adaptation network in Figure \ref{fig:architecture}(c) maps the CLIP embedding of the driving emotion to our proposed emotion latent space. It consists of eight fully-connected (FC) layers with 384 hidden units. ReLU is used as the non-linear activation function in the hidden layers. The input is a 16-dim latent code randomly sampled from a normal distribution. We concatenate the CLIP embedding of the driving emotion to the intermediate embedding after the 4th FC layer.
The dimension of the output emotion feature is set to 896.

\subsection{Additional Training and Testing Details}
We use the Monte-Carlo ELBO loss \cite{portaspeech} to train the variational audio-to-intensity predictor for 9000 steps (about 3 hours). We train our intensity-aware talking-head generation model for about 8 hours on 4 NVIDIA TITAN RTX GPUs. We use Adam optimizer \cite{adam} with $\beta_1 = 0.5 $ and $\beta_2 = 0.999$. The learning rate is set to $1.5 \times 10^{-4}$ for the transformer and $2 \times 10^{-4}$ for other modules. For intensity pseudo-labeling, we find the neutral face of the same person using an emotion classifier \cite{emotionfan}. For precise evaluations during testing, we first crop and align the faces following \cite{atvg} before computing PSNR, SSIM, FID, and F-LMD. Additionally, to assess synchronization confidence, we preprocess the generated videos following PC-AVS \cite{pc-avs}.

\subsection{Additional Analysis on Pseudo-Labeling Method}
\begin{figure}
\begin{center}
\hspace{-0mm}
\includegraphics[width=0.92\columnwidth]{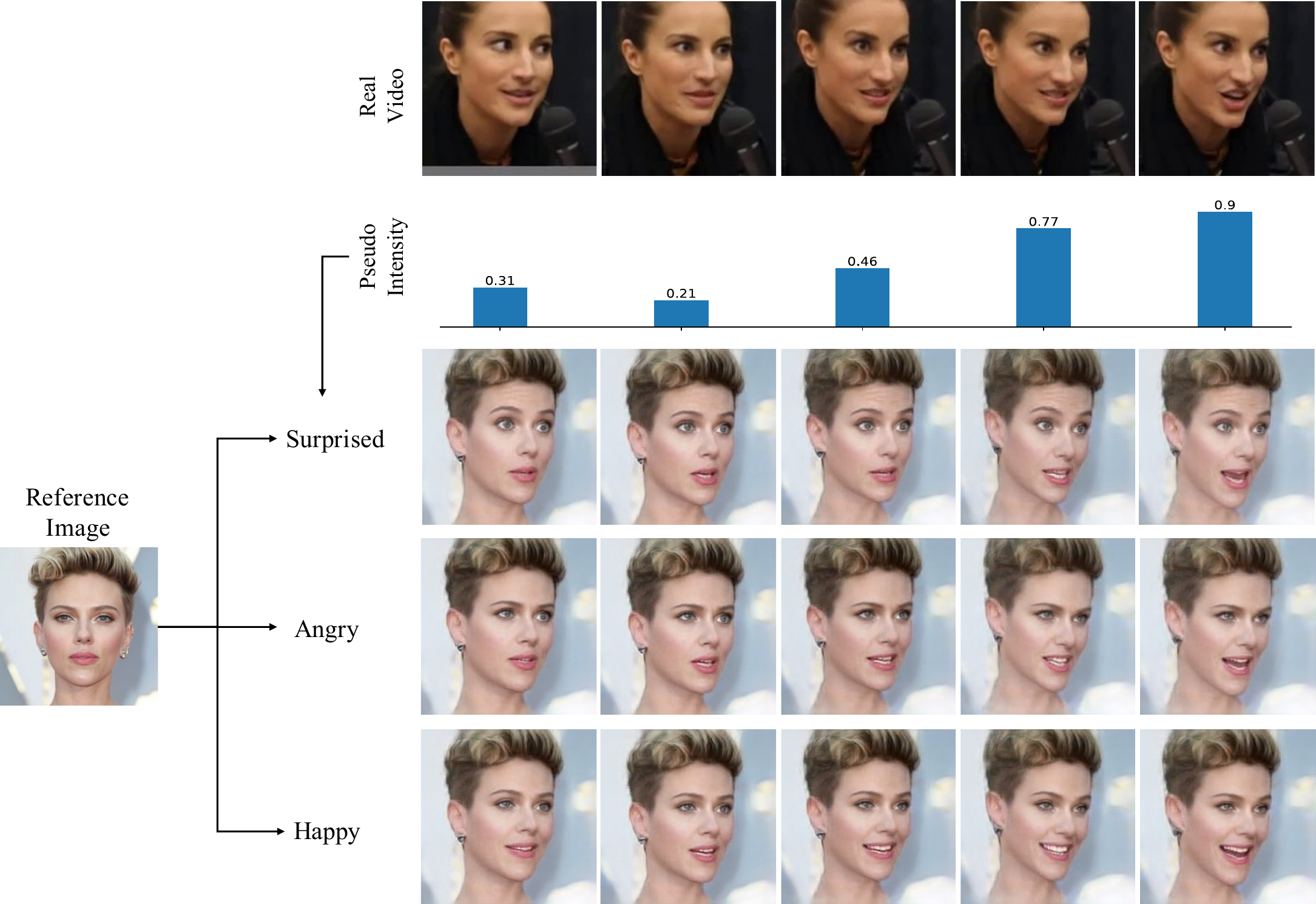} \vspace{1mm}
\caption{ We show a few frames from a real video, 
the inferred pseudo intensity and the corresponding generated video using the inferred intensity. Our pseudo-labeling method accurately captures the emotion intensity from real talking-heads.
} \vspace{-0mm}
\label{fig:transfer2}
\end{center}
\end{figure}

In this section, we present additional qualitative results to validate the correctness of our intensity pseudo-labeling method. 
As shown in Figure \ref{fig:transfer2}, the computed pseudo-intensity reliably reflects the expression intensity of each real frame: the \textit{slightly surprised} face (in the second column) is assigned a value of $0.21$; while the \textit{extremely surprised} face (in the last column) is assigned a value of $0.9$. We further use the pseudo-intensity as the target intensity to generate talking-heads with three different emotions, i.e., \textit{surprised}, \textit{angry}, \textit{happy}.
The generated expression intensity precisely  corresponds to the target intensity across different emotions. 

\subsection{Qualitative Comparisons with Constant Emotion Intensity}
\begin{figure}
\begin{center}
\hspace{-0mm}
\includegraphics[width=\columnwidth]{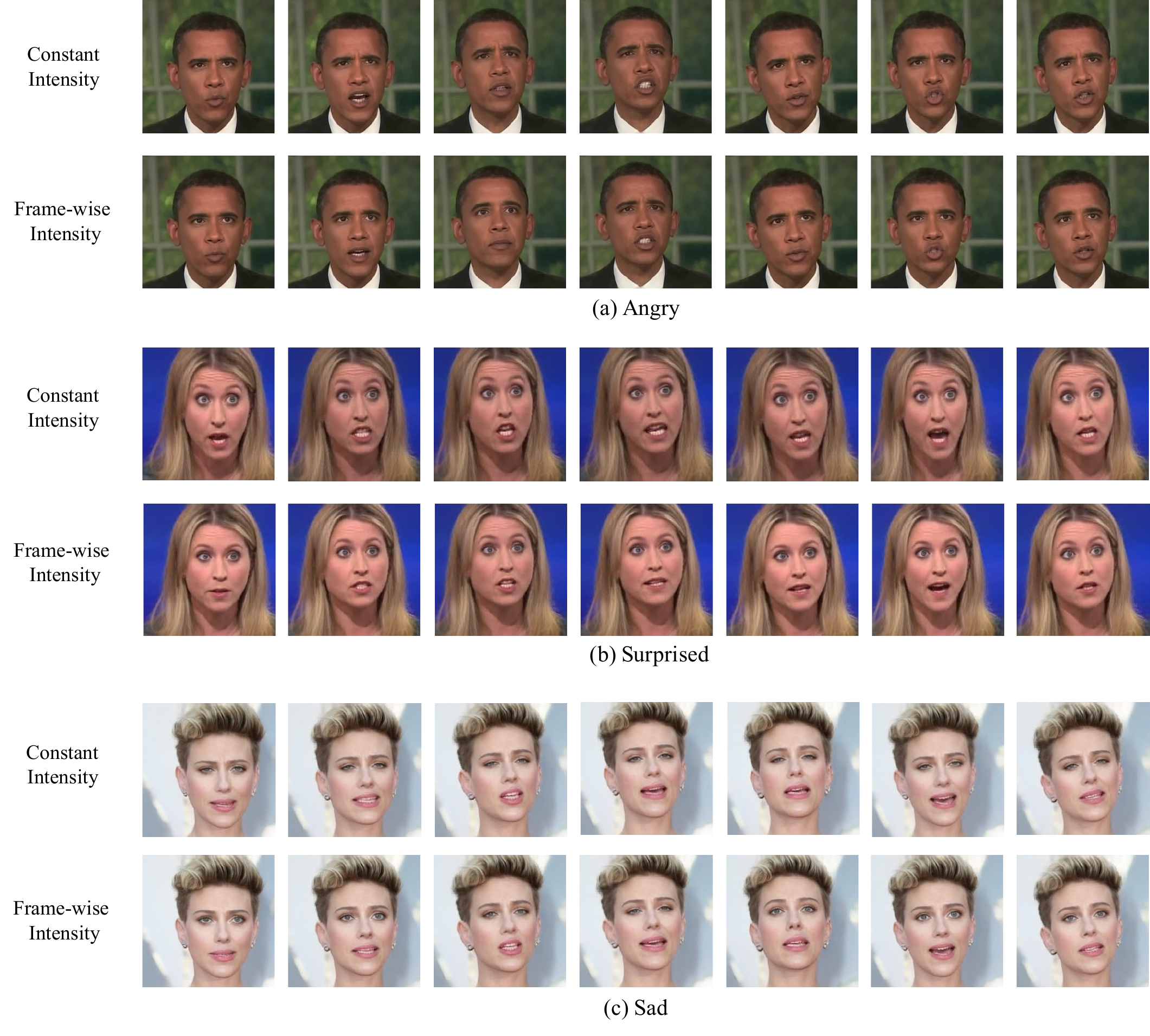} \vspace{1mm}
\caption{Qualitative comparisons between frame-wise intensity and constant intensity for different emotions.
} \vspace{-2mm}
\label{fig:constant1}
\end{center}
\end{figure}

In this section, we provide qualitative comparisons between our method with using constant expression intensity for talking-head generation in Figure \ref{fig:constant1}. As shown in the figure, when expression intensity remains constant throughout the video, it appears artificial and lacking in depth. In contrast, our generated results exhibit smoothly varying expression intensity across frames, resulting in accurate real-life human expressions.

\subsection{Quantitative Results of Emotion Classification}
We present the emotion classification accuracy for each class in Table \ref{tab:emo_classification}. Our method achieves $100\%$ accuracy on \textit{angry}, \textit{neutral}, \textit{surprised} and outperforms the previous state-of-the-art, EAT, by a remarkable margin on \textit{happy}, \textit{disgusted} and \textit{fear}. Our proposed latent space reorganization method effectively rearranges emotion features with varying intensity levels, resulting in a more discriminative latent distribution across classes.  The only class that we observe a slight accuracy drop compared to EAT is \textit{contempt}. For this emotion, we notice many videos where the intensity performed by the actors does not accurately portray the corresponding intensity label, which could negatively affect the learning of the emotion latent space.
\begin{table*}[!ht] 
\centering
\resizebox{\textwidth}{!}{%
  \begin{tabular}{l|ccccccccc}
    \toprule
     & Happy  & Angry & Disgusted & Fear &  Sad & Neutral &  Surprised & Contempt & Average \\
    \midrule
   {Wav2Lip \cite{wav2lip}}  & 0.00 & 25.64 & 0.00 & 0.00 & 0.00 & 91.25 & 0.00 & 0.00 & 17.87   \\
     MakeItTalk  \cite{makeittalk} & 0.00 & 25.64 & 0.00 & 0.00 & 0.00 & 75.00 & 0.00 & 0.00 & 15.23   \\
     AVCT  \cite{avct} &  0.83 & 25.64 & 0.00 & 0.00 & 0.00 & 69.38 & 0.00 & 10.08 & 15.64  \\
    {EAMM \cite{eamm}}  & 23.33 & 84.48  & 9.40 & 0.00 & 0.00 & 98.13 & 94.02 & 72.27 & 49.85  \\
    EAT \cite{EAT} & 84.17 & \textbf{100.00} & 48.72 & 16.52 & 49.17 & \textbf{100.00} & \textbf{100.00} & \textbf{94.96} & 75.43  \\
    \midrule
  Ours & \textbf{99.16} & \textbf{100.00} & \textbf{59.83} & \textbf{37.39} & \textbf{75.00} & \textbf{100.00} & \textbf{100.00} &  {89.92} & \textbf{83.55} \\
    \bottomrule
  \end{tabular}
  } \\ 
 \caption{ Emotion classification accuracy on MEAD \cite{mead}. 
  }\label{tab:emo_classification} \vspace{-0mm}
\end{table*} 

\end{document}